%% file: transversity2024-passek.tex
\documentclass[a4paper,11pt]{article}
\usepackage{pos}

\input{commands-tw3DVMP.tex}

\input{commands-NLOimpact.tex}

\title{Twist-3 contribution to deeply virtual electroproduction of pions}

\author{Kornelija Passek-K.}

\affiliation{
Theoretical Physics Division, Rudjer Bo\v{s}kovi\'{c} Institute \\
HR-10000 Zagreb, Croatia}

\emailAdd{passek@irb.hr}

\abstract{
We discuss deeply virtual meson production (DVMP), 
focusing on the role of higher-twist contributions
in the description of deeply virtual pseudoscalar mesons 
at experimentally accessible energies. 
The standard collinear approach at the lowest twist does not adequately 
describe deeply virtual $\pi_0$ production. 
By incorporating twist-2 transversity generalized parton distributions (GPDs) 
and a twist-3 meson distribution amplitude,
we have determined the twist-3 contribution, which includes both 
the 2-body ($q\bar{q}$)  and 3-body ($q\bar{q}g$) meson Fock components. 
Two methods to regularize the end-point singularities 
are introduced - quark transverse momenta
and a gluon mass.
The resulting cross sections show good agreement with experimental data, 
paving the way for a more comprehensive confrontation of theory 
and experiment at leading order and beyond.
}

\FullConference{7th International Workshop on “Transverse phenomena in hard processes and the transverse structure of the proton (Transversity2024)\\
 3-7 June 2024\\
Trieste, Italy\\}

\begin{document}
\maketitle

\section{Introduction}

The study of high-energy nucleon structure has been primarily
advanced through deeply inelastic scattering (DIS).
The extracted parton distribution functions (PDFs)
represent the probabilities of finding a parton carrying a certain longitudinal momentum fraction of the nucleon momentum,
and thus reveal the nucleon's one-dimensional structure.
Hard-exclusive processes provide insight into the transverse distribution of partons,
and the associated generalized parton distributions (GPDs)
give access to the nucleon’s 3D structure
(see \cite{Belitsky:2005qn,Diehl:2003ny} and references therein).
GPDs are functions of three variables: $x$,
the parton's "average" longitudinal momentum fraction;
$\xi$, the longitudinal momentum transfer (skewness);
and $t$, the momentum transfer squared.
The evolution of GPDs with energy is encapsulated in
the dependence on the factorization scale.
At leading twist-2, there are eight quark GPDs and eight gluon GPDs,
classified by parity and chirality,
as well as distinct GPDs for different quark flavors.
Chirally even GPDs
($H$, $E$, $\tilde{H}$, $\tilde{E}$)
are the most extensively investigated,
while the chiral-odd GPDs, i.e,  transversity (parton helicity flip) GPDs,
($H_T$, $E_T$, $\tilde{H}_T$, $\tilde{E}_T$)
are less well-known.

Hard exclusive processes are successfully described by
the handbag mechanism, in which only one quark from the incoming nucleon
and one from the outgoing nucleon participate in the hard subprocess, while all other partons
remain spectators.
The simplest and most thoroughly investigated process to which
this approach has been applied
is Compton scattering, $\gamma^{(*)} N \rightarrow \gamma N$,
while meson electroproduction, $\gamma^{(*)} N \rightarrow M N'$,
represents a natural extension and offers access to different quark flavors.
A prerequisite for the handbag mechanism is the presence
of at least one large scale, which allows for the use of
a perturbative expansion in the strong coupling constant
and the power, i.e., twist, expansion.
Two kinematic regions have been extensively studied:
the deeply virtual (DV) region, where the virtuality $Q^2$
of the incoming photon is large and the momentum transfer $(-t)$
from the incoming to the outgoing nucleon is small;
and the wide-angle (WA) region, where $(-t)$, $(-u)$, and $s$
are all large, while $Q^2$ is smaller than $(-t)$
(with $Q^2=0$ in the case of photoproduction).
Factorization proofs exist to all orders
for deeply virtual Compton scattering (DVCS) \cite{Collins:1998be}
and deeply virtual meson production (DVMP) \cite{collins},
with the process amplitudes factorizing into hard, perturbatively calculable
subprocess amplitudes and GPDs that encapsulate the soft hadron-parton
transitions and the hadron structure.
General factorization proofs are still lacking for WA processes,
although it has been shown that factorization holds to next-to-leading order
in the strong coupling for wide-angle Compton scattering (WACS) \cite{rad98,DFJK1}
and to leading order for wide-angle meson production (WAMP) \cite{huang00}.
It has been argued that in the symmetric frame, where skewness is zero,
the amplitudes can be represented as a product of subprocess
amplitudes and form factors that represent $1/x$ moments of GPDs at zero skewness.
The DV and WA regions provide complementary access to GPDs
at small and large $(-t)$, respectively.
In this work, we discuss the DV and WA production of pseudoscalar mesons.

The experimental data for deeply virtual pion production
\cite{hermes10,CLAS14,hall-A-2020,COMPASS-2019}
suggest the high significance of transversely polarized photons,
which are not accounted for by the leading twist-2
$\gamma^{*}_L N \rightarrow \pi N'$ contributions.
A twist-3 calculation
has been proposed, which
incorporates twist-2 transversity GPDs
and twist-3 pion corrections.
The calculation, including only the twist-3 2-body
pion Fock component (Wandzura-Wilczek approximation),
has already achieved successful agreement with the data \cite{GK5}.
Similarly, the experimental data for wide-angle pion production
\cite{anderson76,zhu05,kunkel17}
indicate that the twist-2 contributions
\cite{huang00} are not sufficient.
However, unlike in deeply virtual meson production (DVMP), the twist-3 contribution to pion photoproduction was found
to vanish in the commonly used Wandzura-Wilczek approximation \cite{signatures}.
In \cite{KPK18}, both 2-body ($q \bar{q}$)
and 3-body ($q \bar{q} g$)
twist-3 Fock components of $\pi_0$
were considered and successfully fitted to CLAS data \cite{kunkel17}.
This work was extended to the photoproduction of $\eta$ and $\eta'$ mesons
\cite{KPK21a}, and
wide-angle electroproduction of $\pi^{\pm}, \pi^{0}$ \cite{KPK21}.
The latter general analytical results for the
subprocess amplitudes were applied to the deeply virtual (DV) limit ($t=0$),
and thus the complete twist-3 meson contribution%
\footnote{The twist-3 contributions
from the nucleon, i.e., twist-3 GPDs,
are not considered. See footnote 3.}
to DVMP has been obtained.
In this limit,
the 2-body twist-3 contribution
suffers from
an end-point singularity,
which was noted already in \cite{GK5}.

We note that the experimental data for DV pion production
are available for $Q^2 < 10$ GeV$^2$ and moderate $x_B$
\footnote{Skewness $\xi$ is related to Bjorken-$x$ 
by $\xi = x_B / (2 - x_B)$.}.
In contrast, for DV vector meson production,
experimental data for $Q^2 < 100$ GeV$^2$ and small $x_B$
exist and show the importance of longitudinal photon contributions,
which can be well described by twist-2 contributions \cite{Cuic:2023mki}.
We remark that the twist-3 effect advocated in \cite{GK5}
also occurs in the exclusive electroproduction of longitudinally polarized vector mesons,
but it is a small effect visible only in some of the spin-density
matrix elements \cite{GK7}.

We report here on the analysis of DV$\pi_0$ production
\cite{Duplancic:2023xrt}.
We use the information on the twist-3 pion distribution amplitude (DA), 
which was obtained from the wide-angle analysis \cite{KPK18}.
Two methods are employed to handle the end-point singularities.
The first is the modified perturbative approach (MPA),
in which the end-point singularity is regularized
by allowing for quark transverse momenta inside the meson,
while the emission and reabsorption of quarks
from the nucleon are still treated collinearly to the nucleon momenta.
In the second approach, we use the standard collinear framework but introduce
a dynamically generated mass in the gluon propagators,
which reflects the fact that gluons, as carriers of strong interactions,
strongly influence the nonlinear dynamics of the infrared sector of QCD.

\section{Subprocess amplitudes at twist-3}

\begin{figure}
\centering
  \includegraphics[width=0.45\columnwidth]{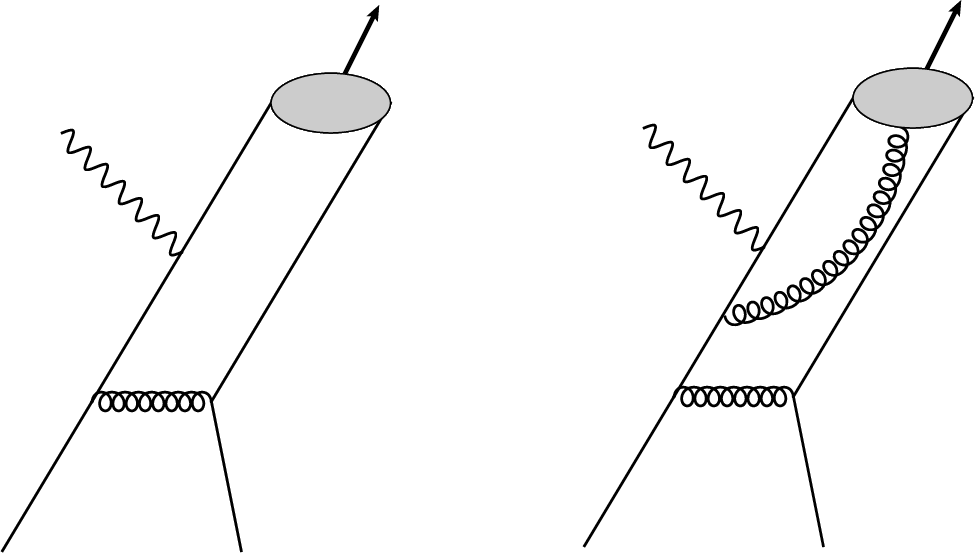} 
\caption{Generic diagrams for 2- and 3-body subprocess amplitudes.}
\label{fig:H}
\end{figure}

The amplitudes ${\cal H}$
corresponding to the subprocesses
$\gamma^{(*)} q \rightarrow \pi q'$
are calculated using handbag diagrams
like those depicted in Fig. \ref{fig:H},
with $\pi$ replaced by the appropriate
2- or 3-body Fock state.
In \cite{KPK21}, the general leading-order (LO) expressions for any
energy range ($s$, $u$, $t$, and $Q^2$) have been obtained.

The 2-body projector $\pi \rightarrow q\bar{q}$
contributes to the subprocess amplitudes
corresponding to the diagrams
depicted on the left in Fig. \ref{fig:H},
and its structure
is given by
\begin{equation}
{\cal P}_{2}^{\pi} \sim f_\pi
   \,\Big\{\gamma_5 \,p\sla \, \phiDA_\pi(\tau,\mu_F) 
                   + \mu_\pi(\mu_F) \, \gamma_5 \Big[\,\phiDA_{\pi p}(\tau,\mu_F)  
- [ \ldots ] \phiDA'_{\pi\sigma}(\tau,\mu_F)
+ [ \ldots ] \phiDA_{\pi\sigma}(\tau,\mu_F)\Big] \Big\}
\, .
\label{eq:2proj}
\end{equation}
The first term in \req{eq:2proj} corresponds to the twist-2 part,
while the twist-3 part is proportional to the chiral condensate
$\mu_\pi = m_\pi^2 / (m_u + m_d) \cong 2$ GeV (at the factorization scale $\mu_F = 2$ GeV).
This parameter is large, and although the twist-3
cross-section for pion electroproduction is
suppressed by $\mu_\pi^2 / Q^2$
compared to the twist-2 cross-section,
for the range of $Q^2$ accessible in current
experiments, the suppression factor is of order unity%
\footnote{
Twist-3 effects arising from twist-3 GPDs lead to additional
power corrections and are expected to be smaller; hence, they are neglected.
}%
.
The 3-body $\pi \rightarrow q \bar{q} g$ projector
\begin{equation} 
{\cal P}_{3}^{\pi} \sim
             f_{3\pi}(\mu_F) \gamma_5 [ \ldots ]\,
  \phiDA_{3\pi}(\tau_1, \tau_2, \tau_g, \mu_F)
\, .
\label{eq:3proj}
\end{equation}
contributes to the subprocess amplitudes corresponding to
the diagram on the right in Fig. \ref{fig:H}.
Here, $\tau_i$ denote the longitudinal momentum fractions of the pion's constituents.
The helicity non-flip amplitudes are generated by twist-2, while
the helicity flip ones are of twist-3 origin.

In addition to the twist-2 DA $\phiDA_\pi$, there are two 2-body twist-3 DAs,
$\phiDA_{\pi p}$ and $\phiDA_{\pi \sigma}$, as well as the 3-body twist-3 DA
$\phiDA_{3 \pi}$.
Twist-3 DAs are related by equations of motion (EOMs).
Using the EOMs and the symmetry properties of the DAs,
it is possible to express the twist-3 subprocess amplitudes
in terms of only two twist-3 DAs, thereby combining 2- and 3-body contributions.
Applying the EOMs also leads to an inhomogeneous linear first-order differential
equation, which is used to determine $\phi_{\pi p}$ (and $\phi_{\pi \sigma}$)
from a known 3-body DA $\phiDA_{3 \pi}$%
\footnote{It is important to note that the same gauge must be used consistently
for the constituent gluon in both the $q\bar{q}g$ projector and the EOMs.}
.
We use the ansatz advocated in \cite{braun-filyanov}
and include also the evolution.

The general structure of the twist-3 subprocess amplitudes
is given by
\begin{eqnarray}
{\cal H}^{\pi,\text{tw3}}
\label{eq:Htw3}
& = & {\cal H}^{\pi,\text{tw3},q\bar{q}} + {\cal H}^{\pi,\text{tw3},q\bar{q}g}
\nonumber \\
& = &
   \big({\cal H}^{\pi,\phi_{\pi p}} + \underbrace{{\cal H}^{\pi,\phi_{\pi}^{\text{EOM}}} \big)
   + \big( {\cal H}^{\pi,q\bar{q}g, C_F}} + {\cal H}^{\pi,q\bar{q}g, C_G} \big)
\nonumber \\
& = &
   {\cal H}^{\pi,\phi_{\pi p}} 
   + \qquad \quad {\cal H}^{\pi,\phi_{3\pi},C_F} \quad \qquad
   + {\cal H}^{\pi,\phi_{3\pi},C_G}
\, ,
\end{eqnarray}
where ${\cal H}^{\pi,\text{tw3},q\bar{q}}$ is the twist-3 2-body contribution
proportional to the $C_F$ color factor, and ${\cal H}^{\pi,\text{tw3},q\bar{q}g}$
is the twist-3 3-body contribution with $C_F$ and $C_G$ proportional parts.
The $C_G$ part is gauge invariant, whereas for $C_F$ contributions,
only the sum of the 2-body and 3-body parts is gauge invariant with respect to the choice
of photon or virtual gluon gauge.
EOMs are used to obtain this sum, as well as the complete twist-3 contribution,
which is expressed in terms of only two twist-3 DAs: $\phi_{3\pi}$ and $\phi_{\pi p}$.
The twist-3 subprocess amplitude for longitudinal photons
vanishes both for photoproduction and DVMP.
Moreover, for photoproduction,
${\cal H}^{\pi,\phi_{\pi p}} = 0$
\cite{KPK18}.
For DVMP, ${\cal H}^{\pi,\phi_{\pi 2}^{\text{EOM}}} = 0$,
and while no end-point singularities are present for $t \neq 0$,
they must be considered in the limit $t \to 0$.
Namely, for DV$\pi$ production,
${\cal H}^{\pi,\phi_{\pi p}} \sim \int_0^1 \frac{d\tau}{\tau} \phi_{\pi p}(\tau)$,
with $\phi_{\pi p}(\tau)$ nonvanishing at the end-points $\tau = 0, 1$.
This singularity requires regularization, for which we
propose two methods below: the introduction of quark transverse momenta
and a gluon mass.


Following the approach in \cite{GK5}, where the WW approximation was used, 
we have calculated the subprocess amplitudes within the MPA, 
taking into account the transverse momenta of the partons entering the pion. 
In this scenario, the DAs of the collinear approximation are replaced by light-cone wave functions. 
The parton transverse momenta are accompanied by gluon radiation. The gluon radiation, as described in \cite{botts89}, has been calculated in the form of a Sudakov factor up to next-to-leading-log 
approximation using resummation techniques and the renormalization group.
The resummation of the logarithms involved in the Sudakov factor can only be efficiently performed 
in the impact parameter space, which we then switch to in this approach. 
The interplay between the quark transverse momenta and the Sudakov factor 
effectively regularizes the endpoint singularity in the 2-body contribution.

A practical disadvantage of the MPA is the large computing time needed 
for the involved multidimensional integrations 
and the demanding calculation of NLO corrections. 
Thus, we have proposed the collinear approach and the regularization 
of the endpoint singularity by the introduction of a dynamically generated gluon mass 
into the gluon propagators. 
The gluon mass generation, which is based on the Schwinger mechanism, 
is currently the subject of intensive studies in the QCD context 
and is motivated by recent evidence for such a phenomenon 
from lattice simulations---see, e.g., \cite{Aguilar:2014tka, Shuryak:2020ktq}.

\section{Numerical results}

\begin{figure}[t]
\begin{center}
  \includegraphics[width=0.5\tw]{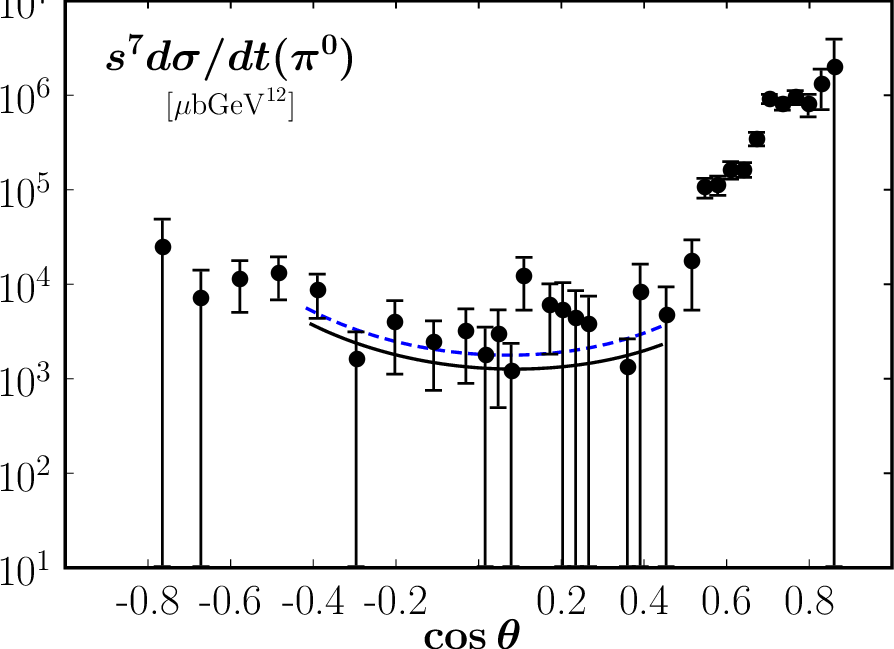}
\end{center}
\caption{
The cross section of $\pi^0$ photoproduction 
versus the cosine of the scattering angle 
in the center-of-mass system
at $s=11.06\,\gev^2$. 
The data are taken from \cite{kunkel17}.
The dashed line represents the fitted results obtained
using the DA parameters from \cite{KPK18} (set KPK). 
The solid line represents the modified DA
parameters from \cite{Duplancic:2023xrt}.
}
\label{fig:wa-data}  
\end{figure}
\begin{figure}
\begin{center}
\includegraphics[width=0.36\tw]{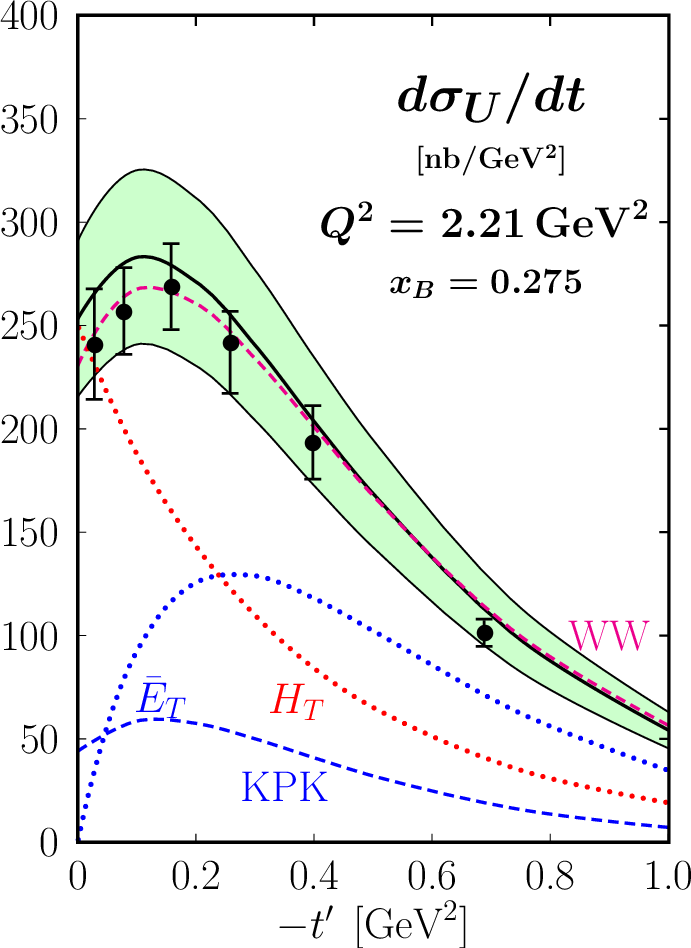} \hspace*{0.05\tw}
\includegraphics[width=0.36\tw]{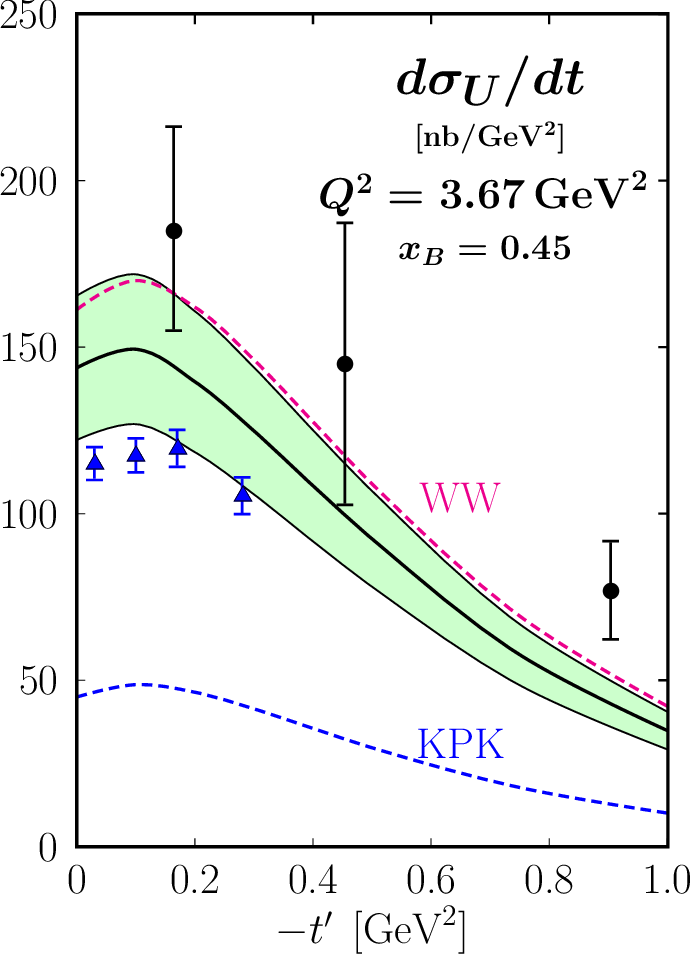}
\includegraphics[width=0.36\tw]{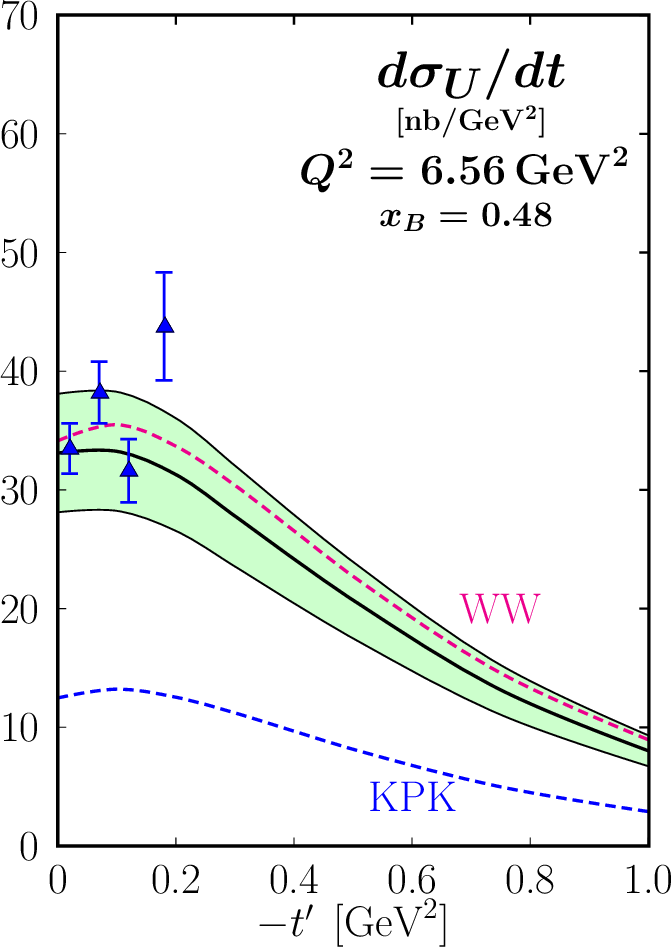}  \hspace*{0.05\tw}
\includegraphics[width=0.36\tw]{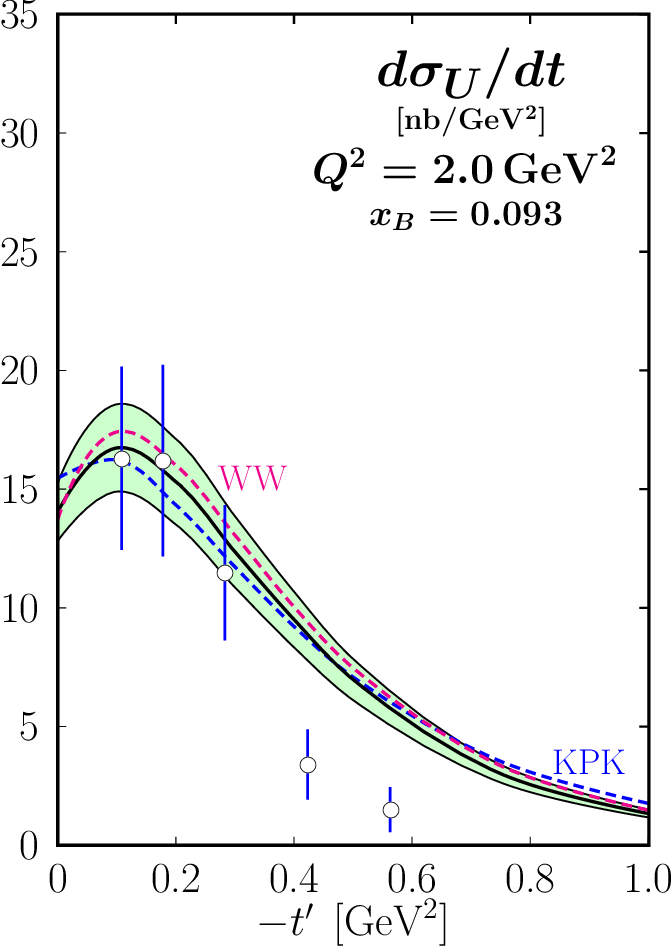}
\end{center}
  \caption{The unseparated cross section versus $t'$ for various
    kinematical settings obtained using MPA. 
The separate contributions from $H_T$ and $\bar{E}_T$ are shown as dotted lines 
for $Q^2=2.21\,\gev^2$. 
The dashed lines are evaluated using the \da{} parameters from \cite{KPK18} (KPK)
and the WW approximation \cite{GK6} (WW).  
The solid lines with error bands,
evaluated from the uncertainties of the GPDs and $\mu_\pi$,
are obtained using modified \da{} parameters 
explained on  Fig. \ref{fig:wa-data}.
The data are taken from CLAS \ci{CLAS14} (full circles), 
Hall A \ci{hall-A-2020} (triangles) 
and COMPASS \ci{COMPASS-2019} (open circles). 
The Hall A data in the upper right plot are at the adjacent kinematics $Q^2=3.57\,\gev^2$ 
and $x_B=0.36$.
}
  \label{fig:T}
\end{figure}
\begin{figure}
\begin{center}
\includegraphics[scale=0.7]{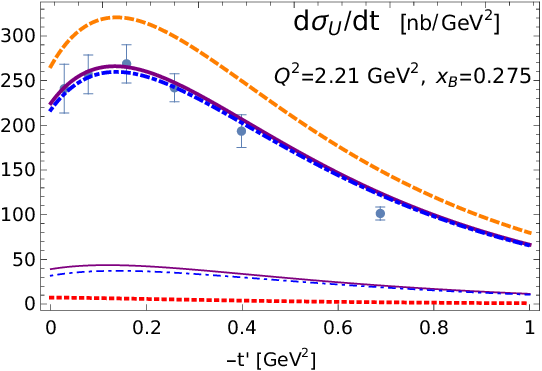}
\includegraphics[scale=0.7]{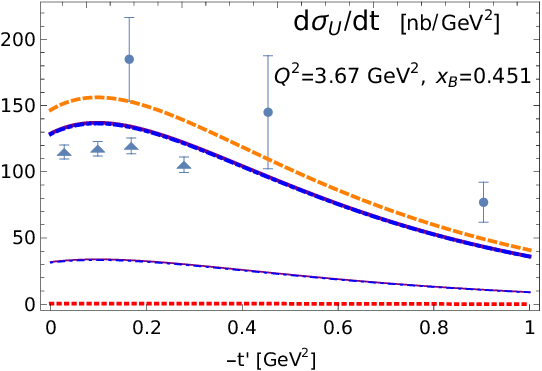}
\\
\includegraphics[scale=0.7]{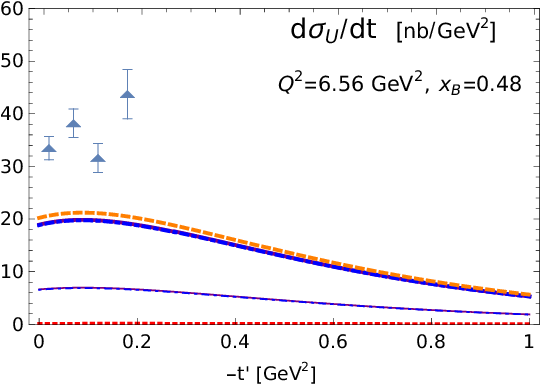}
\includegraphics[scale=0.7]{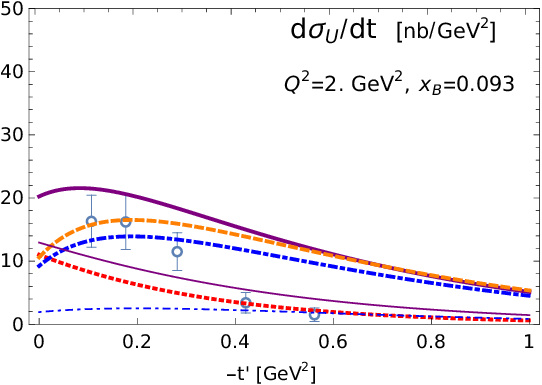}
\caption{The unseparated cross section 
 versus $t'$
for various kinematical settings,
obtained using 
the collinear approach with the gluon mass
$m_g^2(Q^2)$ \cite{Aguilar:2014tka}
The thin lines denote 
the cross sections obtained using the 
pion DA parameter set from \cite{KPK18}
and the thick lines the ones obtained
using the modified set explained
on Fig. \ref{fig:wa-data}.
The specific contributions are denoted:
solid line $d\sigma_U/dt$, 
dot-dashed line $d\sigma_T/dt$, 
and
dotted line $d\sigma_L/dt$. 
The dashed line represents the 
$d\sigma_U/dt$ WW contribution.
The experimental data
are marked as on Fig.
\ref{fig:T}. 
}
\label{fig:sigmaU-coll} 
\end{center}
\end{figure} 

The helicity amplitudes 
for the process 
$\gamma^* N \to \pi N'$, are
given by convolutions of the GPDs and subprocess amplitudes
${\cal H}^{\pi}$.
The amplitudes describing the contribution of
longitudinal photons 
are driven by twist-2 subprocess amplitudes and the
GPDs $\tilde{H}$ and $\tilde{E}$ (parton helicity non-flip).
On the other hand, contributions from transverse photons 
are described
by the above-discussed twist-3 subprocess amplitudes (parton helicity flip)
and the transversity GPDs.
Here we take into account $H_T$ and $\bar{E}_T = 2 \tilde{H}_T + E_T$.
Moreover, considering the energy region of
available experimental data, we restrict our investigation
to valence quark GPDs.

The total differential cross section is then expressed through
the longitudinal component $d\sigma_L/dt$
(depending on $\tilde{H}$, $\tilde{E}$),
the transverse component $d\sigma_T/dt$
(dominantly influenced by $H_T$ and $\bar{E}_T$), 
and interference terms
$d\sigma_{TT}/dt$ ($\bar{E}_T$) and
$d\sigma_{LT}/dt$ ($\tilde{E}$, $H_T$).
The data for both interference terms are plagued by large errors,
but show the dominant contribution of transverse part since $d \sigma_{TT}$
tends to be large and $d\sigma_{LT}$ small.
Furthermore, the available data mostly provide the unseparated cross section defined as
$d\sigma_U/dt = d\sigma_T/dt + \epsilon d\sigma_L/dt$,
where $\epsilon$ is the ratio of the longitudinal and transversal polarization of
the virtual photon.

We have used the GPDs proposed in \cite{GK6,GK7,GK3}.
In these works, the GPDs are constructed from the zero-skewness GPDs,
and their products with appropriate weight functions are treated
as double distributions from which the skewness dependence
of the GPDs is generated \cite{musatov99}.
The forward limit of the GPD $H_T$ is given by the transversity parton density, while
a similar parametrization is used for $\bar{E}_T$.
For the described parameterization of the zero-skewness GPDs, combined with
a suitable weight function,
the double-distribution integral can be performed analytically \cite{GK3}.
There are five parameters for each GPD and flavor.
As an initial comparison with experimental data, we have used parameters from
\cite{GK6,GK7}.
Since the subprocess amplitudes now include 3-body contributions as well,
we have tested modifications of the transversity GPDs, while leaving the
refitting for future work.
However, there are constraints on these GPDs from lattice QCD \cite{LQCD,LQCD2}
and we test those in our analysis.

In this work we demonstrate that the twist-3 contribution,
combined with a moderate adjustment of the transversity GPDs,
leads to reasonable results for a sample of
kinematic settings $(Q^2, x_B)$ where experimental data
are available from CLAS \cite{CLAS14},
the Hall A collaboration \cite{hall-A-2020}, or COMPASS \cite{COMPASS-2019}.
We emphasize that our analysis focuses
on $\pi^0$ production, as in this case, contributions
from transverse photons, which are of twist-3 origin,
are dominant.
This contrasts with charged pion production,
where longitudinal photons play a decisive
role, particularly at small $-t'$.
This dominance is primarily due to the significant contribution from the pion pole \cite{GK5}.

The twist-3 pion DAs $\phi_{3\pi}$ and $\phi_{\pi p}$ are connected by EOMs 
and expressed through the same parameters \cite{KPK18}. 
Two parameters are taken from a QCD sum rule analysis, while one is fixed 
in \cite{KPK18} by a fit to the WA photoproduction data \cite{kunkel17}. 
The photoproduction cross section obtained using that DA parameter set 
is represented in Fig.~\ref{fig:wa-data} by a dashed line. 
The modified DA parameter set, needed for better agreement with DVMP data, 
produces the cross section denoted by the solid line. 
It is within errors and also describes well the photoproduction data.

In Fig.\ \ref{fig:T} we show the MPA results for the unseparated cross section. 
The results evaluated using the DA parameter set from \cite{KPK18}, denoted by KPK, 
are below the data. 
We have introduced the modified set (solid line), which describes both photoproduction 
(see Fig. \ref{fig:wa-data}) and DV data well. For comparison, 
we also show the results obtained with the WW approximation \cite{GK6}. 
The contribution of the transverse $d\sigma_T/dt$ cross-section is dominant for all displayed 
kinematical settings. 
The smallness of the predicted longitudinal cross section at an $x_B$ of about 0.3-0.4 
is in agreement with a few experiments where Rosenbluth separation was performed 
\cite{defurne,mazouz}. 
However, at the COMPASS kinematics, i.e., for smaller $x_B$, 
$d\sigma_L/dt$ is substantially larger. It amounts to about $40\%$ of the transverse cross section.

Similar findings are obtained using the collinear approach 
with the dynamical gluon mass as a regulator in the 2-body twist-3 contribution. 
The corresponding predictions for $d\sigma_U/dt$ are displayed in Fig.\ \ref{fig:sigmaU-coll}. 
Here separate longitudinal and transverse cross-sections are depicted. 
One can clearly see that the longitudinal cross section $d\sigma_L/dt$ (dotted line) 
is much smaller than $d\sigma_T/dt$ (dot-dashed line) 
for relatively low $Q^2$ and $W$ at which CLAS 
and Hall A data are available (first three figures). 
In contrast, in the low $x_B$ kinematics, as for COMPASS data (lower right figure), 
the longitudinal cross section cannot be neglected and is of comparable 
size as the transverse one. 
In that energy region the NLO twist-2 corrections 
should be included. We note that in this work we have 
aimed to present a proof of concept and offer an insight into the interplay of contributions 
for the collinear approach. The $Q^2$ dependence of the results should be improved 
and a more comprehensive collinear analysis involving fits performed.

\section{Conclusions}

We studied the twist-3 contributions to DVMP beyond the WW approximation. 
The 3-body twist-3 contributions are important for ensuring gauge invariance. 
They are smaller than the 2-body contributions, 
but the 3-body DA modifies the 2-body twist-3 distribution amplitudes and, as a result, 
affects the 2-body twist-3 contributions through the equations of motion. 
The 3-body twist-3 DA was fixed by the adjustment to the wide-angle pion 
photoproduction data. 
The end-point singularity present in the 2-body twist-3 contribution 
was regularized using MPA, 
in which quark transverse momenta and Sudakov suppressions are taken into account. 
As a second regularization method, we proposed the use of a dynamically generated gluon mass 
and a collinear approach, which offers easier inclusion of NLO corrections.
The improved twist-3 analysis shows that twist-3 contributions dominate in DV$\pi^0$ production 
at accessible energies, except for the COMPASS kinematics where $x_B$ is small. 
The NLO corrections to twist-2 may be important 
in the COMPASS kinematics.
We stress that wide-angle meson production is similarly dominated by twist-3 contributions 
and provides information on the pion DA as well as complementary information on GPDs at large $-t$.
The information about the small and large $-t$ behavior of GPDs
is valuable for the study of the 3D partonic structure of the proton.

\end{document}

%% file: commands-tw3DVMP.tex
\newcommand{\be}{\begin{equation}}
\newcommand{\ee}{\end{equation}}
\newcommand{\ba}{\begin{eqnarray}}
\newcommand{\ea}{\end{eqnarray}}
\newcommand{\ci}[1]{\cite{#1}}
\newcommand{\tw}{\textwidth}                          

\newcommand{\da}{{DA}}

\newcommand{\sla}{\hspace*{-0.20cm}/}

\def\={\,=\,}

\def\gev{\,{\rm GeV}}

\def\Li{\relax\ifmmode{\text{Li}_{2}}\else{Li$_2${ }}\fi}

\def\phiDA{\phi}

%% file: commands-NLOimpact.tex
\usepackage{hepunits}
\usepackage{hepnames}
\usepackage{xpatch}
\makeatletter
\xpatchcmd\@HepConStyle
 {\edef\@upcode{\updefault}}
 {\ifdefined\shapedefault\edef\@upcode{\shapedefault}\else\edef\@upcode{\updefault}\fi}
 {}{}
\makeatother

\newcommand{\req}[1]{(\ref{#1})}
